\documentclass[prd,nofootinbib,preprintnumbers,preprint]{revtex4}
\usepackage{indentfirst}

\usepackage{dcolumn}
\usepackage{bm}
\usepackage{indentfirst}
\usepackage{bm}
\usepackage{mathrsfs}
\usepackage{comment}
\usepackage{amsfonts}
\usepackage{amssymb}
\usepackage{dsfont}
\usepackage{euscript}
\usepackage{graphicx}
\usepackage{wrapfig}
\usepackage{mathtools}
\usepackage{pifont}
\usepackage[dvipsnames]{xcolor}
\usepackage{tensor}
\usepackage{amsmath,latexsym,esint}
\usepackage{cmap}
\usepackage[figurename=Figure,font=footnotesize]{caption}
\usepackage{subcaption}
\usepackage{hyperref}
\usepackage{float}
\usepackage{xcolor}
\usepackage[toc,page]{appendix}

\begin{document}
	\preprint{}
	\title{Charged nutty black holes are hairy}
	\author{Dmitri Gal'tsov}
	\email{galtsov@phys.msu.ru}
	\author{Rostom Karsanov}
	\email{karsanovrz@my.msu.ru}
	\affiliation{Faculty of Physics, Moscow State University, 119899, Moscow, Russia
	}
	
	\begin{abstract}
    We uncover the physical nature of the electric and magnetic monopoles discovered by McGuire and Ruffini on Misner strings accompanying charged nutty black holes, showing that these strings carry singular, nonuniform flows of electric and magnetic fields. These fields inevitably have nonzero divergence, thereby simulating the effective electric and magnetic charge densities along the strings. The latter create a complex short-range electromagnetic hair zone around the horizon, making the combined Misner-Dirac strings classically observable. Typical features of this new type of hair are presented. We also note that rotation can act as a hair generator even in the absence of NUT.	    
	\end{abstract}
	
	\maketitle
	
	\section{Introduction}
    For many years, solutions of general relativity with the Newman-Unti-Tamburino (NUT) parameter \cite{Newman:1963yy,Misner:1963fr,brill} were considered unphysical due to the presence of Misner strings and the chronological horizon around them. However, it was recently discovered   \cite{Clement:2015cxa,Clement:2015aka,Clement:2016mll} they they are less pathological in the Bonnor interpretation \cite{Bonnor:1969ala,Sackfield71}, which views Misner strings as physical singularities without imposing time periodicity condition.
Indeed, geodesics pass freely through the Misner strings and the horizon, chronology violation does not occur for geodesic \cite{Clement:2015cxa} and non-geodesic \cite{Clement:2016mll} observers, and reasonable thermodynamics can be developed \cite{Hennigar:2019ive,BallonBordo:2019vrn,Bordo:2019slw,Durka:2019ajz,Wu:2019pzr}. For a review of attitudes toward nutty black holes, see \cite{Bordo:2019slw}.
    
    Here we show that by extending Bonnor's interpretation to {\em Dirac's} strings, another paradox of nutty solutions can be resolved.  
    In Einstein-Maxwell and/or supergravity theories, charged black holes are current-free solutions of Maxwell's equations, a phenomenon called "charge without charge" \cite{Misner:1957mt}. However, addition of NUT parameter  makes the situation unclear in light of the discovery by McGuire and Ruffini \cite{Mcguire:1975wq} of additional electric and magnetic charges on the Misner-Dirac (MD) strings. The existence of such charges was later noticed in many papers \cite{Clement:2017otx,Clement:2019ghi, BallonBordo:2020mcs,Abbasvandi:2021nyv,Awad:2022jgn}, but their physical nature remained unclear. Here, we suggest an explanation and show that these solutions can be considered as hairy black holes.
	 
	Dirac strings, accompanying $U(1)$ magnetic monopoles  in flat space \cite{Dirac:1931kp},
mimic infinitely thin semi-solenoids carrying a uniform magnetic flux. This can be seen by calculating the Maxwell two-form corresponding to the magnetic monopole potential $A=-p \cos\theta d\varphi$ in  local Cartesian coordinates $x,\;y$
near the polar axis 
\cite{Clement:2022pjr}: $\sin\theta=u=\sqrt{x^2+y^2},x=u\cos\varphi,\,y=u\sin\varphi$, so that
\begin{equation*}
	  \label{dphi}
		d\varphi=u^{-2}(xdy-ydx)=\partial_{x}\ln u \;dy-\partial_y\ln u \; dx.
	\end{equation*} Obviously, the one-form $d\varphi$ is singular on the polar axis $u=0$ and is not closed in the sense of distributions:
\begin{equation}\label{ddphi}
dd\varphi=\left(\partial_x^2+\partial_y^2\right)\ln u\,dx\wedge dy=2\pi\,\delta^2(\mathbf{x})\,dx\wedge dy,
	\end{equation}
   where $\delta^2(\mathbf{x})=\delta(x)\,\delta(y)$.
	It follows that the Maxwell two-form $F=dA$  in addition to the usual term $F_{\theta\varphi}=p\sin\theta$ has a singular part representing two delta-like  opposite field fluxes which can be attributed to the North  and South  Dirac strings:
\begin{equation}\label{singFflat}
		{\cal F}=\mp 2\pi p \,\delta^2(\mathbf{x})\;dx\wedge dy.
	\end{equation}
	 
	An effective magnetic monopole was observed at the tip of a magnetized nanoscopic ferromagnetic needle, \cite{Beche:2013wua}, realizing a design of a split solenoid.
However, the semisolenoids themselves are not classically observable, although they can be detected by Aharonov-Bohm scattering \cite{Aharonov:1959fk} if the quantization condition is not imposed. With this condition, Dirac strings are fictitious \cite{Lipkin:1982ts}.
	
	The goal of this paper is to show that, in contrast, the MD strings in charged nutty  solutions are classically observable via the peculiar "hair" they create around the black hole. The reason lies in the nonuniformity of the singular electromagnetic fluxes carried by Dirac strings in presence of NUT: gravity causes them to decay with distance, thereby pushing some field lines of force (LF) into the bulk. The latter partially terminate at the horizon, partially on the strings themselves, and partially extend to infinity.  
	The arising short range field structure  literally resembles ``hair'' around the black hole  \cite{Volkov:1998cc}. This hair exists despite the linearity of the Maxwell field, the required non-linearity  being provided by gravity. 
    
	\section{Static nutty dyons}
    We start with the Brill solution  
	\begin{align} \label{metricB}
		&ds^2=-\frac{\Delta}{\Sigma}\left(dt+2n (\cos\theta+c) d\varphi\right)^2+\frac{\Sigma}{\Delta}dr^2+\Sigma d\Omega^2,\\ 
        &\Delta=r^2-2mr+q^2+p^2-n^2,\quad \Sigma=r^2+n^2\nonumber
	\end{align}
     where $m$ is the mass, $n$ is the NUT parameter, $q$ and $p$ are electric and magnetic charges. The metric has Misner string singularities\cite{Misner:1963fr}: north at $\theta=0$  and south $\theta=\pi$, whose configuration is determined by the parameter $c$. For $c=\pm 1$ there is only south (north) string, for $c=0$ both strings are present.  There is no singularity at $r=0$, so the overcharged $m^2+n^2<q^2+p^2$ solutions represent wormholes \cite{Clement:2015aka}. 
	
	With the potential one-form
	\begin{equation}
		\label{potential}
		A=-\frac{qr-np}{\Sigma}(dt+2n(\cos\theta +c)d\varphi)-p\cos\theta d\varphi,
	\end{equation}
	the metric \eqref{metricB} is a solution of the Einstein-Maxwell equations everywhere except for the polar axis $\theta=0,\pi$ where the Ricci tensor has a string-like singularity \cite{Clement:2022pjr}. It should be noted that the vector potential is sensitive to the value of $c$. Therefore, the formal coordinate transformation $t\to t+2n(c-c')\varphi$, which leads to a change in $c\to c'$ in the metric, also modifies the electromagnetic potential, the modification of which is not reduced to the choice of gauge. So, the choice of $c$ is not merely the choice of gauge, but a physical choice. The Komar angular momentum integral over the infinite sphere also depends on $c$ and diverges unless one chooses $c=0$ \cite{Clement:2019ghi}. So in what follows we set $c=0$.

    Applying Eq. \eqref{ddphi} to \eqref{potential}, we  find that the Maxwell tensor also has  a singular hidden magnetic flux:  
	\begin{equation}\label{singF}
		{\cal F}=\mp
		2\pi\delta^2(\mathbf{x})\left(p+2n(qr-np)/\Sigma\right)dx\wedge dy,\end{equation}
	which coincides with 
	\eqref{singFflat} for $n=0$.
	But for $n\neq0$
	the magnetic field  depends on the coordinate $r$,  gradually decaying with   distance. Such a field has a nonzero divergence, which is manifest as some  magnetic current:
\begin{equation*}\label{Bianchi 1}
		\nabla_\mu \tilde{\cal F}^{\nu \mu}=4\pi J^\nu_m.
	\end{equation*}
    
	But \eqref{singF} is not the full singular part of the Maxwell tensor. The electric-magnetic   duality of the Einstein-Maxwell equations implies that electric fluxes must also be present. These can be read off using the potential $B$, which generates the dual Maxwell tensor  $\tilde{F}=dB$:
	\begin{equation}\label{potentialDual}
		B=-\frac{pr+nq}{\Sigma}(dt+2n\cos\theta d\varphi)+q\cos\theta d\varphi. 
	\end{equation}
	The singular part of the dual Maxwell tensor is
	\begin{equation}\label{singFd}
		\tilde{{\cal F}}=\mp
		2\pi\delta^2(\mathbf{x})\left(-q+2n(pr+nq)/\Sigma\right)dx\wedge dy.\end{equation}
	It generates a non-zero singular electric current via
	\begin{equation*}\label{Max1}
		\nabla_\mu  {\cal F}^{\nu \mu}=4\pi J^\nu_e.
	\end{equation*}
	A straightforward calculation now gives:
	\begin{align} 
		&J^\mu_e=\delta^\mu_0 \rho_e \delta^2(\mathbf{x})/\Sigma, \quad J^\mu_m=\delta^\mu_0 \rho_m \delta^2(\mathbf{x})/\Sigma,\nonumber \\\label{Scharges_e}
		&\rho_e= {n\left(p(r^2-n^2)+2rnq\right)}(r^2+n^2)^{-2},\\\label{Scharges_m}
		&\rho_m=- {n\left(q(r^2-n^2)-2rnp\right)} (r^2+n^2)^{-2}.
	\end{align}
	The densities  \eqref{Scharges_e} and \eqref{Scharges_m} are related by  duality  \begin{equation}\label{EMduality}
	    \rho_m=\rho_e(q\to p,p\to-q).
	\end{equation} 
    As follows from our derivation, these densities do not describe true physical charges, but are the consequence of the non-uniformity of fluxes  \eqref{singF} and \eqref{singFd}.   In other words, the effective solenoids now are split at every point, mimicking a continuous distribution of monopoles (and similarly electric charges). It should be noted that the constant terms $p$ and $q$ in \eqref{singF} and \eqref{singFd} do not enter into these effective charge densities and can be omitted (bare Dirac strings), but the quantities proportional to $n$ are observable (Misner-Dirac strings).  
	
	\section{Bulk fields}
    
    Now we turn to the bulk description. The bulk fields can be found either from \eqref{potential} or \eqref{potentialDual}: 
\begin{align}\label{Fbulk}
		&F=\Phi dr\wedge (dt+2n\cos \theta d\varphi)+\tilde{\Phi}\Sigma\sin\theta d\theta\wedge d\varphi,\\ &\Phi=\Sigma^{-2}\left[q(r^2-n^2)-2 npr\right],\nonumber
	\end{align}
    \vspace{-.2cm}
	 \noindent and $\tilde{\Phi}$ is   $\Phi$ with the replacement $q\to p,\,p\to -q$. 

\begin{figure}[H]
	\centering
\includegraphics[width=0.5\linewidth]{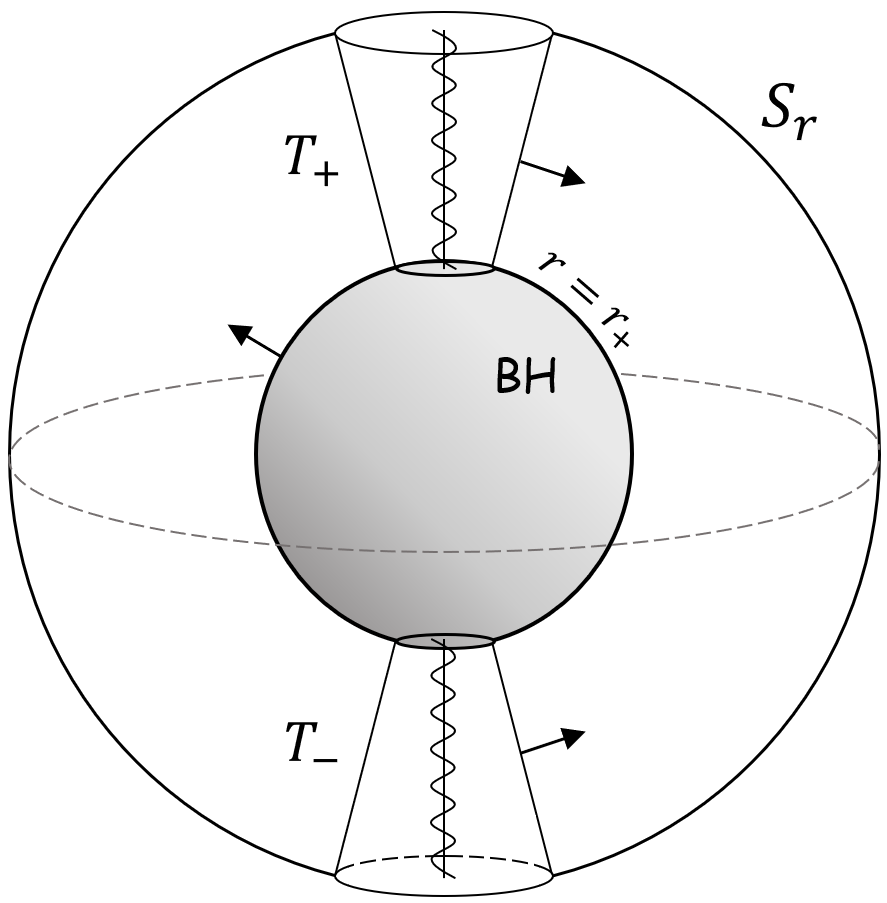}
	\caption{Domain of integration for fluxes of electric and magnetic fields. (The arrows indicate the directions of the corresponding normal vectors.)}
	\label{BHcones}
\end{figure}

     Consider a regular region between the horizon and the  sphere $S_r$ of an arbitrary radius $r$ cutting the segments of Misner strings by conical surfaces \cite{Barbagallo:2025tjr}, as shown in Fig.\eqref{BHcones}. 
     By the Gauss theorem,  the effective charges inside $S_r$ can be obtained by integrating the corresponding fluxes as follows
    \begin{align}
		&Q(r) =\frac{1}{4\pi}\int \limits_{S_r}\!\!F^{tr}\sqrt{-g} d\theta d\varphi =q-\frac{2n(pr+nq)}{r^2+n^2}\,\label{Eflux  through sphere}\\
		&P(r)=\frac{1}{4\pi}\int \limits_{S_r}\!\!\tilde{F}^{tr}\sqrt{-g}d\theta d\varphi  =p+\frac{2n(qr-np)}{r^2+n^2}.\label{Mflux  through sphere}
	\end{align}
	When $r\to\infty$, we get $P_\infty=p$
	and $Q_\infty=q$, but at finite $r$ the  effective charges are different, indicating that the bulk fields are not radial. 
    
    The integrals \eqref{Eflux  through sphere},\eqref{Mflux  through sphere}  over  the horizon  $Q_H=Q(r_+),\; P_H=P(r_+)$ can be expressed through the effective horizon charge densities 
\begin{equation}\label{HorizonDensities}
    \begin{aligned}
        Q_H=\int \limits_H \!\!\rho^H_e\sqrt{\gamma_H}  d\theta d\varphi, \qquad\;
        P_H=\int \limits_H \!\!\rho^H_m \sqrt{\gamma_H}  d\theta d\varphi,
    \end{aligned}
\end{equation}
    where $\gamma_H$ stands for the spatial section of the horizon. 

	Similarly, the effective charges located on the North and South segments of the strings inside $S_r$ are given by fluxes through the surrounding conical surfaces:  
\begin{equation}
		Q_\pm=\pm \frac{1}{4\pi}\int \limits_{T_\pm} \!\!F^{t\theta}\sqrt{-g}drd\varphi ,\qquad
		P_\pm=\pm  \frac{1}{4\pi}\int \limits_{T_\pm}\!\!\tilde{F}^{t\theta}\sqrt{-g} drd\varphi.
\end{equation}

     The crucial point is that the charges obtained from the bulk viewpoint are given precisely to the integrals from the linear densities  \eqref{Scharges_e}, \eqref{Scharges_m} obtained  from the {\em divergencies of the singular fluxes}:
\begin{equation}\label{strings charges}
		Q_+=Q_-=\int \limits_{r_+}^r\rho_e dr, \;\;\;\qquad P_+=P_-=\int \limits_{r_+}^r\rho_m dr.    
	\end{equation}
For any distance $r$, the balance equation $Q(r)=Q_H+Q_++Q_-$ holds, similarly for   $P(r)$. 

The charges \eqref{Eflux  through sphere} and  \eqref{Mflux  through sphere} have {\em transition points} $r=r_{e,m}$ where $Q(r_e)=0$ and $P(r_m)=0$  and change signs, which are also transition points of the   string  densities, but now  $\rho_m(r_e)=0$ and  $\rho_e(r_m)=0$. Thus, there are zones of opposite  charge densities between which arise confined LFs  (SS-hair).  Similarly, confined LFs exist between the  horizon and the string segments  (SH-hair).

The Ricci tensor of the Brill solution also has a distributional part, which has been found to be responsible for the violation of the null-energy condition in the wormhole sector \cite{Clement:2022pjr}. To apply our method, we note that for a metric independent of $t$, the relevant Christoffel symbols are antisymmetric: $\Gamma_{\mu \nu t}=(\partial_\nu g_{\mu t}-\partial_\mu g_{\nu t})/2$, which leads to the form description: 
\begin{equation*}
    \mathbf{\Gamma}= \Gamma_{\mu \nu t}dx^\mu \wedge dx^\nu=-d\Omega, \qquad \Omega=g_{tt}dt+g_{t\varphi}d\varphi.
\end{equation*}
Then, using \eqref{ddphi}, we can find the singular values
\begin{equation*}
    \Gamma_{xyt}=-\Gamma_{yxt}=\pm 2\pi  n\Delta\Sigma^{-1}\delta^2(\mathbf{x})\mathcal.
\end{equation*}
Finally, raising the index in $R_{at}=g^{\mu \nu}\partial_\mu \Gamma_{\nu a t}$, we get
\begin{equation*}
 \sqrt{-g}   R^a_t= \pm2\pi n \Delta\Sigma^{-2}\epsilon^{ab}\partial_b \delta^2(\mathbf{x}),
\end{equation*}
that coincides with the result of \cite{Clement:2022pjr}.

\section{Rotation}\vspace{-.2cm} 
    For the Kerr-Newman-NUT solution (in the notation of \cite{Clement:2022pjr})
	  we obtain: 
	\begin{align}\label{Kerr strings}
\rho^\pm_e=n\frac{2qr(n\pm a)+p(r^2-(n\pm a)^2)}{(r^2+(n\pm a)^2)^2}, 
	\end{align}
     and $\rho_m^\pm$ is given by\eqref{EMduality}. For the  horizon  
 \eqref{HorizonDensities} with $\gamma_H=(r_+^2+a^2+n^2)\sin \theta$ one has 
 
	 \begin{equation}\label{kerr charge distr on horizon}
	\rho^H_e=\frac{q[r_+^2-(n+a\cos \theta)^2]-2pr_+(n+a\cos \theta)}{4\pi(r_+^2+(n+a\cos \theta)^2)^2}, 
\end{equation} 
     and $\rho_m^H$, obtained using \eqref{EMduality} (for $p=n=0$ this coincides with Eq. (3.113c) of  \cite{Thorne:1986iy}. 
    Now the horizon can have sectors of opposite signs, connected by LFs, forming HH-hairs, even at $n=0$. Thus, rotation also acts as a hair generator.
    
	\section{ Hair patterns}
    We define electric and magnetic LFs following Christodoulou and Ruffini \cite{Christodoulou:1973} (see also \cite{Mcguire:1975wq,Hanni:1973fn}) as the integral curves of the field measured locally by a non-rotating observer at rest whose four-velocity is     
$u_\mu=(-\sqrt{1/|g^{tt}|},0,0,0).$
Such electric and magnetic fields read 
\begin{align*}
    E^\mu=u_tF^{t\mu},\;\qquad
H^\mu=u_t\tilde{F}^{t\mu}.
\end{align*}
Since we also have $E^\varphi=H^\varphi=0$, we may choose any plane $\varphi={\rm const}$.
In the paper \cite{Hanni:1973fn} it was shown that this definition is equivalent to the lines of force defined as the locus of points with a fixed value of field flux $\Phi$, having the slope of the lines given by the formula
\begin{equation*}
	\frac{dr}{d\theta}=-\frac{\partial \Phi/\partial \theta}{\partial \Phi/\partial r}=\frac{\sqrt{-g}F^{tr}}{\sqrt{-g}F^{t\theta}}.
\end{equation*} 

In addition, we compactify the space using the radial variable $\arctan(r/k), k=\rm{const}$, which brings infinity to a finite point $\pi/2$, shown by the limiting circle. (We note inevitable disproportions due to this map.)
	 
For a given set of parameters, we present  pair plots of {\it electric} LFs (left panel) and {\it magnetic LFs} (right panel). The red (blue) color represents positive (negative) signs of effective charge densities. Accordingly, LFs are outgoing (incoming) there.  The horizon is a boundary of the gray disk in the center, the signs of horizon charges $Q_H$ and $P_H$ being shown with the same color conventions. The green lines represent the transition spheres $r_e$ and $r_m$ .  

\begin{figure}[H]
		\centering		\includegraphics[width=0.7\textwidth]{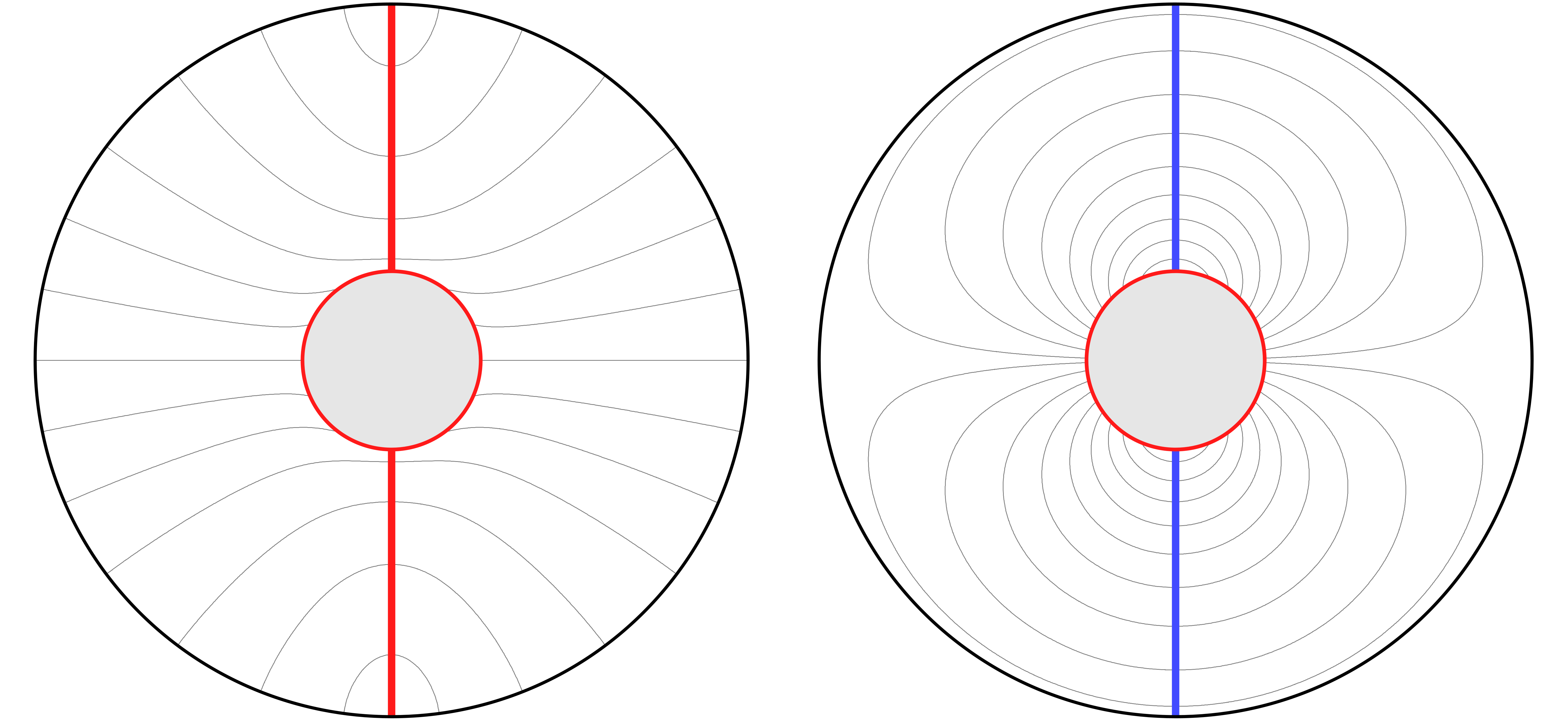}
		\caption{ $r_e<r_+$ ( $m=0.75,\; n=1.2,\; q=0.5,\; p=0$). Electric LFs  start at the positively charged MD strings and at the horizon, spreading directly to infinity.  Magnetic LFs are confined, starting form the horizon and closing on the  MD strings (SH-hair).} 
		\label{static 1}
	\end{figure}

\begin{figure}[H]
		\centering
		\includegraphics[width=0.7 \textwidth]{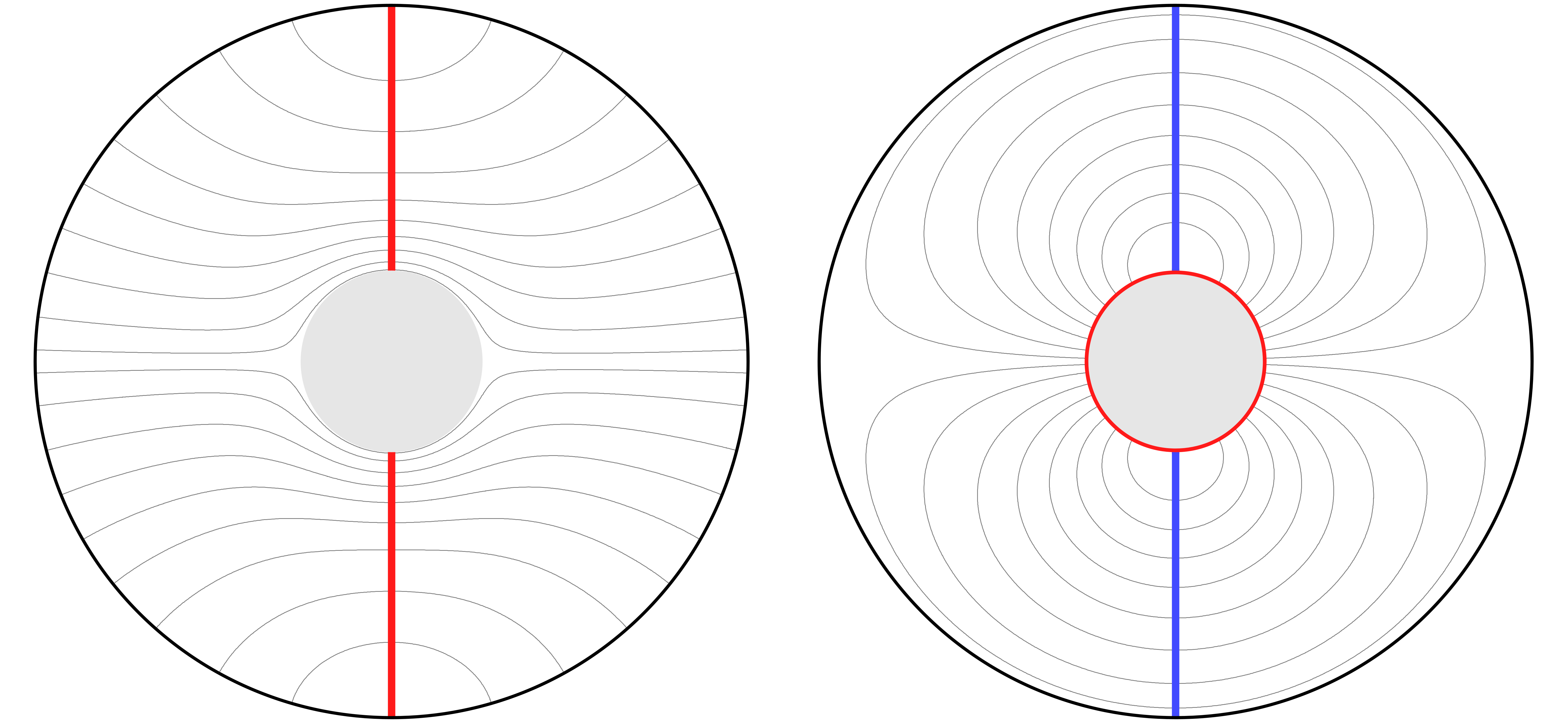}
		\caption{$r_e=r_+$  ($m=1,\; n=2,\; q=2,\; p=0$). Electric LFs propagate from the MD strings to infinity and not touching the horizon. Magnetic field is again of the SH hair type}
		\label{static 2}
	\end{figure}

\begin{figure}[H]
		\centering
		\includegraphics[width=0.7\textwidth]{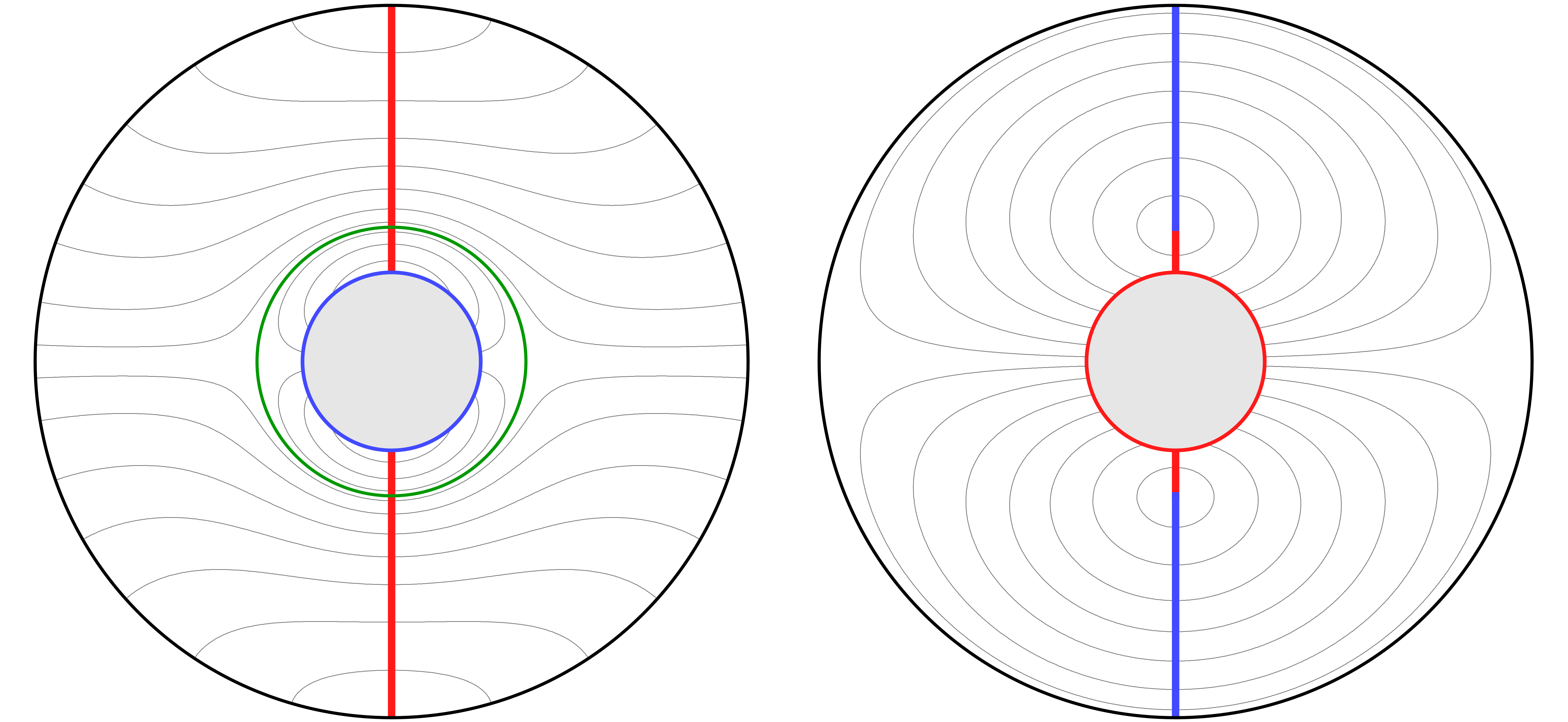}
		\caption{$r_e>r_+$ ($m=0.75,\; n=1.3,\; q=1.5,\; p=0$). The green circle $r=r_e$ separates the confined SH electric hair sector from the zone of LFs  spreading  to infinity. Magnetic hair now has both SH and SS sectors.}
		\label{static 3}
	\end{figure}

\begin{figure}[H]
		\centering
\includegraphics[width=0.7\textwidth]{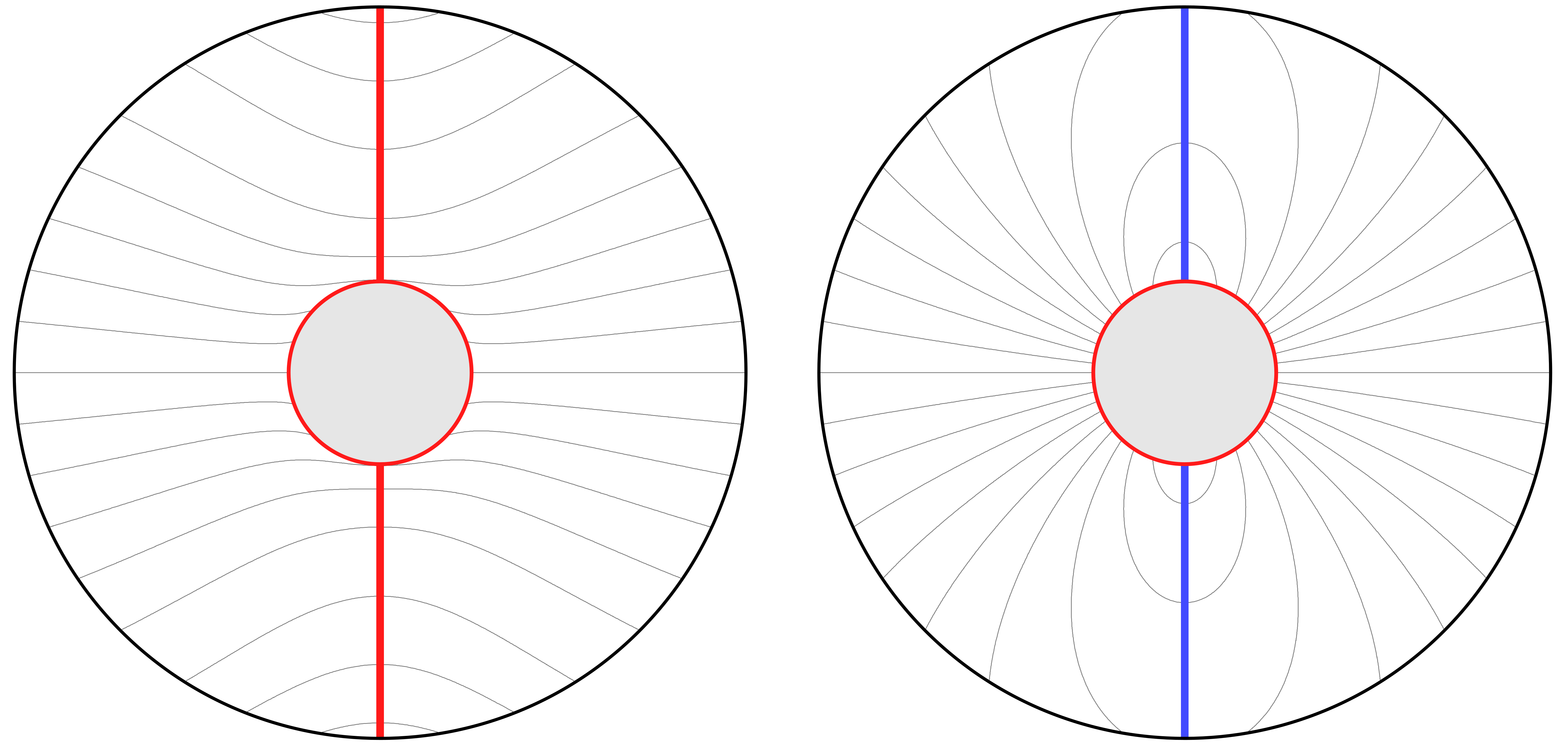 }
		\caption{$r_m<r_e<r_+$ ($m=2,\; n=1,\; q= p=0.71$).   All the electric LFs starting either on the horizon or on the strings spread to infinity, while part of the magnetic LFs starting on the horizon  end on Misner strings forming SH hair.}
		\label{BH dyon q=p rh>re rh>rm}
	\end{figure}

\begin{figure}[H]
		\centering
		\includegraphics[width=0.7\textwidth]{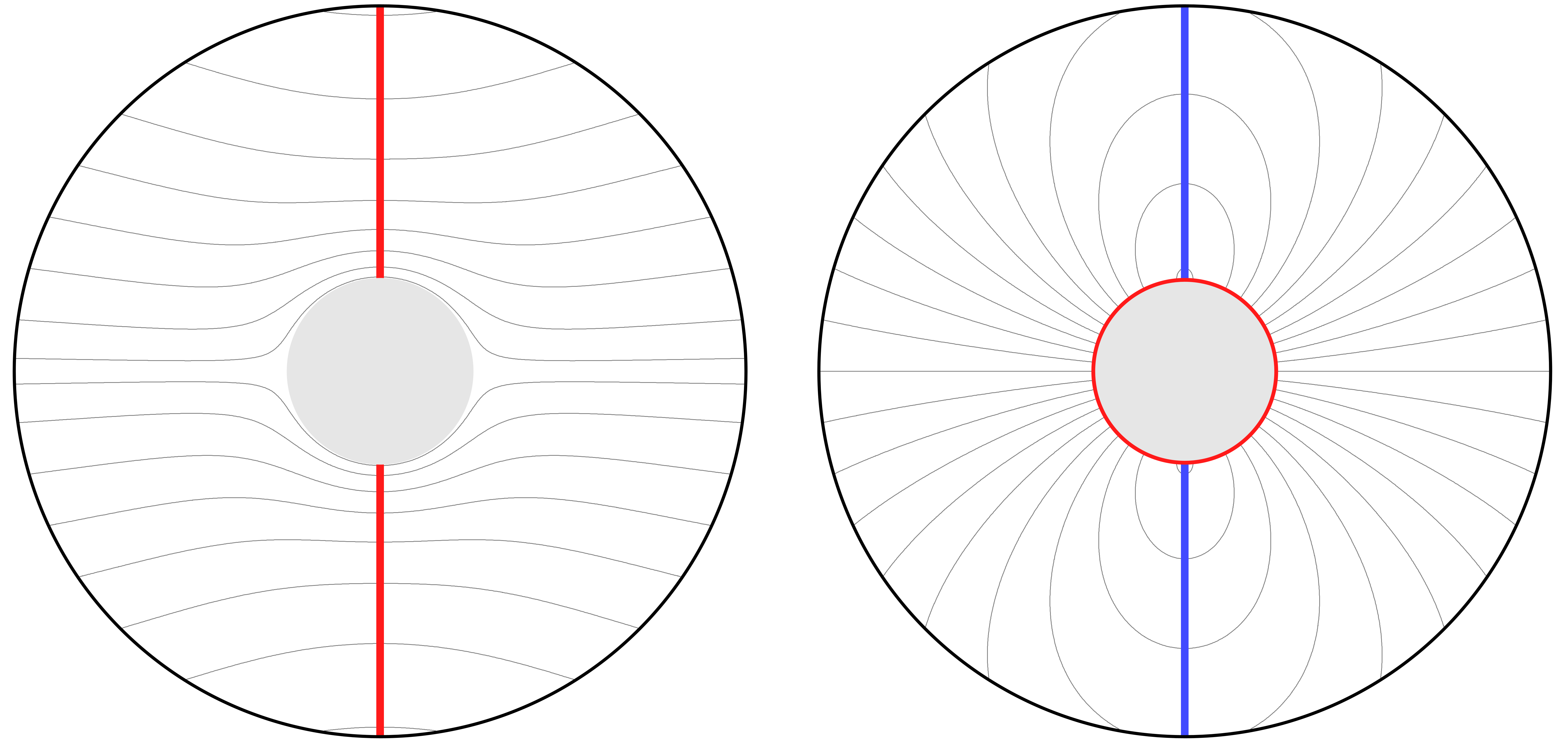 }
		\caption{$r_m<r_+,\; r_e=r_+$ ($m\approx1.21,\;n=1,\; q= p=0.71$). The electric flux throughout the horizon vanishes, so the electric charge at infinity is due to   the LFs, coming from the Misner strings only.}
		\label{BH dyon q=p rh=re rh>rm}
	\end{figure}

\begin{figure}[H]
		\centering
		\includegraphics[width=0.7\textwidth]{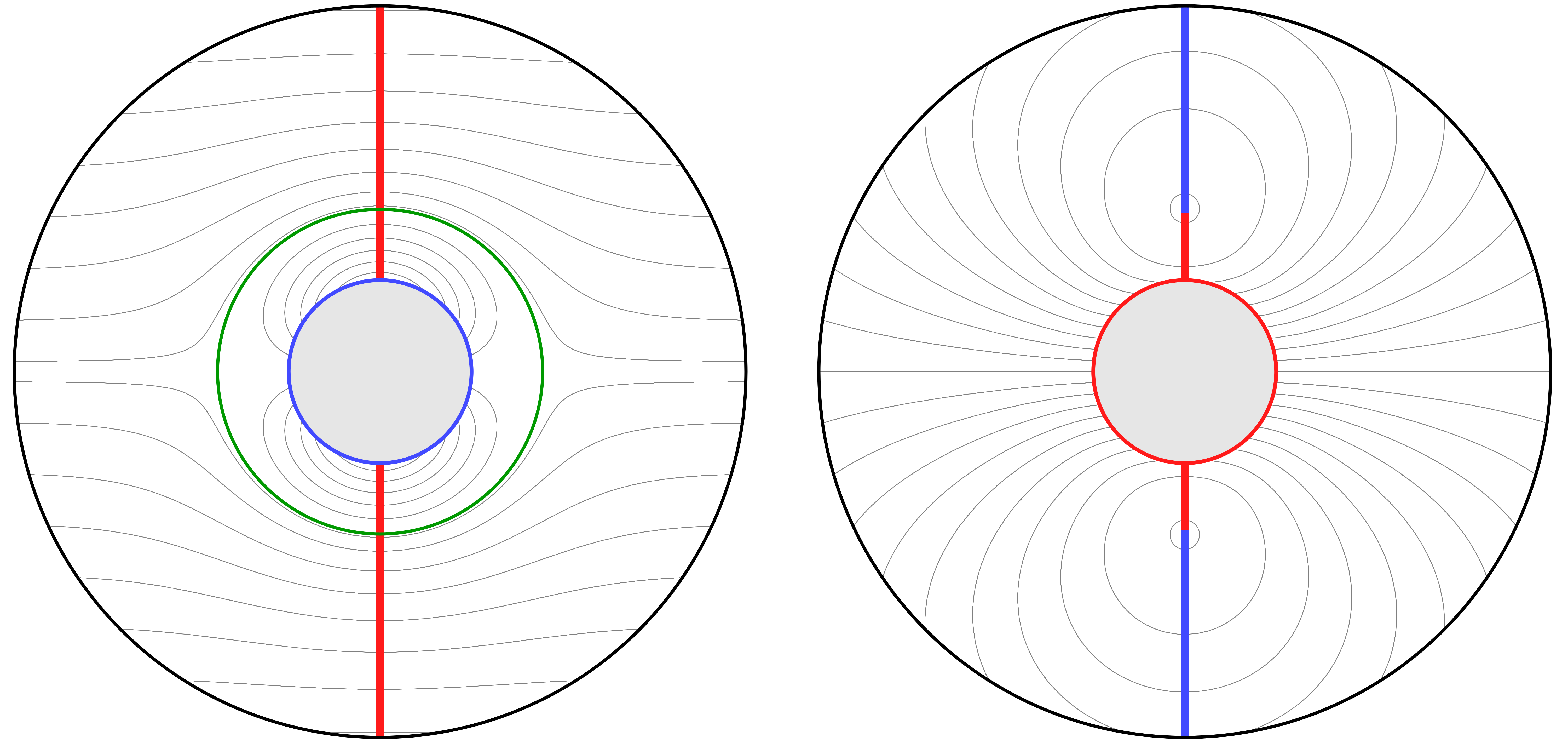 }
		\caption{$r_m<r_+<r_e$ ($m=0.6,\; n=1,\; q=p=0.71$). Part of electric lines are confined between the horizon and strings (SH-hair), and part of magnetic lines starts and ends on the strings (SS-hair). }
		\label{BH dyon q=p rh<re rh>rm}
	\end{figure}

\begin{figure}[H]
		\centering
		\includegraphics[width=0.7\textwidth]{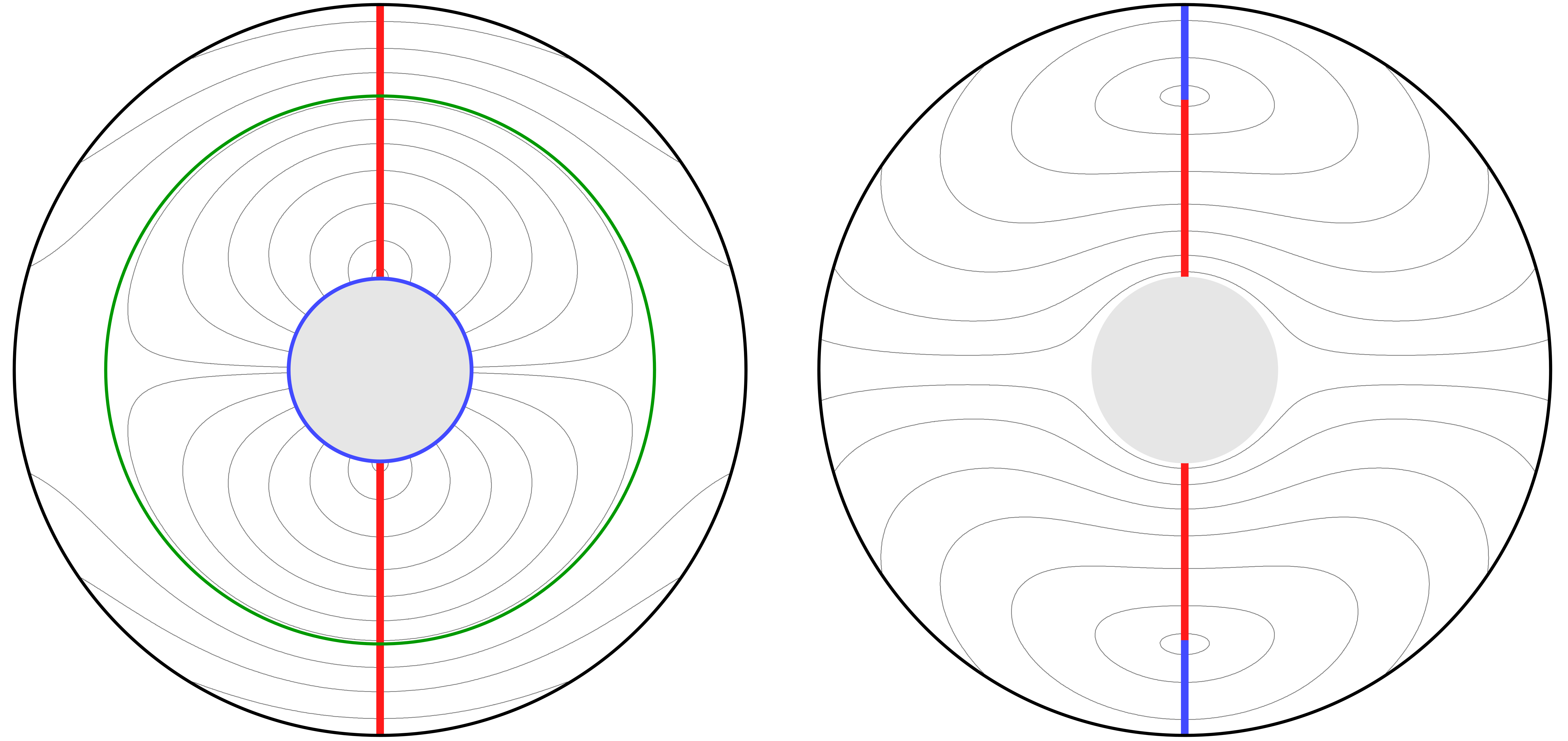 }
		\caption{$r_m=r_+,\; r_e>r_+$($m\approx 0.21,\;n=1,\; q=p=0.71$). Electric lines are partly confined (SH-hair), magnetic LFs do not touch the horizon, but contains SS-hair sector.}
		\label{BH dyon q=p rh<re rh=rm}
	\end{figure}

\begin{figure}[H]
		\centering
		\includegraphics[width=0.7\textwidth]{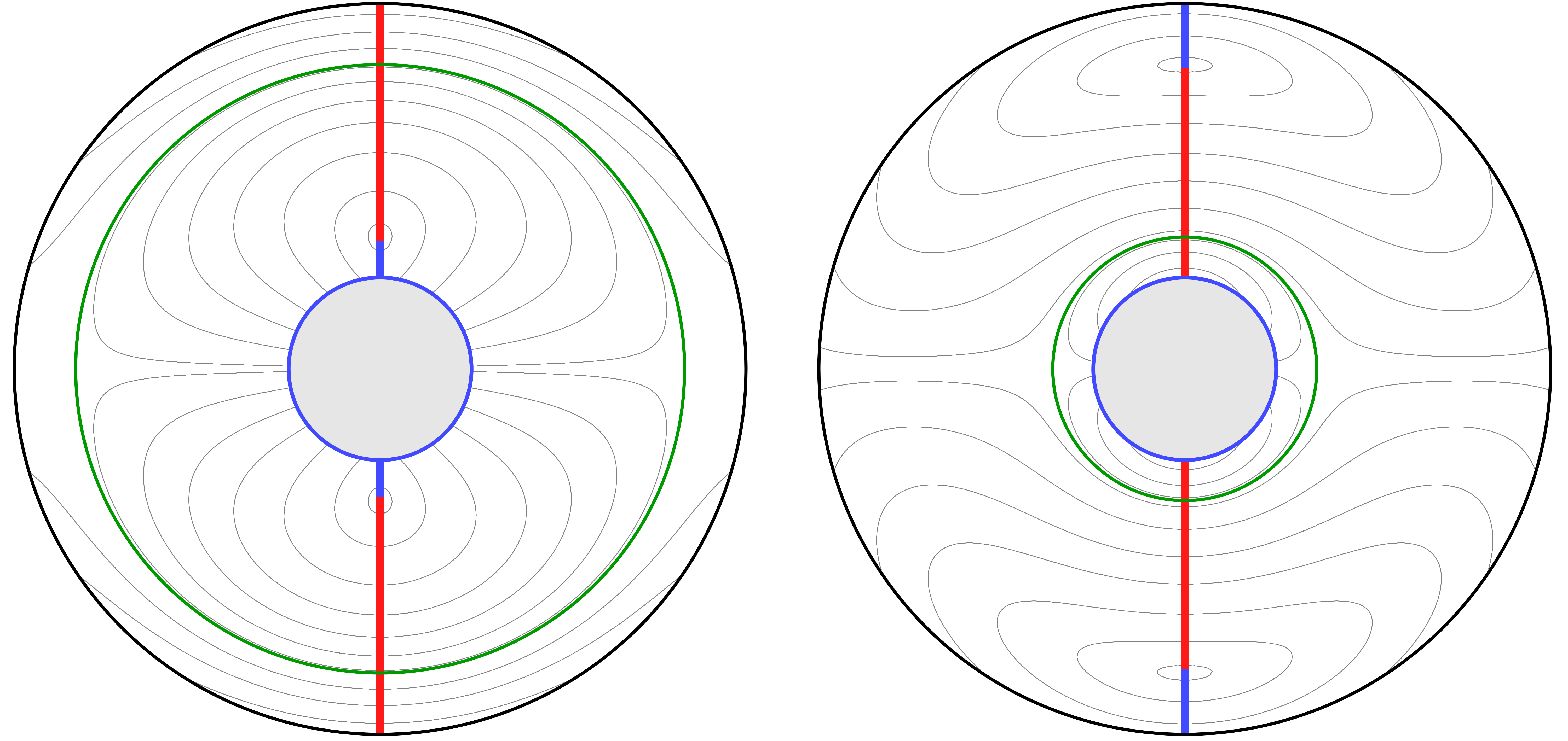 }
		\caption{$r_+<r_m<r_e$ ($m=0.15,\; n=1,\; q=p=0.71$).  For both fields the MD charge densities change signs, so there are confined SH and SS hair.}
		\label{BH dyon q=p rh<re rh<rm}
	\end{figure}

{\bf   Single charge.}
	Let's start with the purely electrical case, where the purely magnetic case is achievable thanks to duality \eqref{EMduality}.  
   Three different possibilities are presented in Fig.\ref{static 1}-\ref{static 3}.
	 \vspace{-.5cm}
	 {\bf Static symmetric dyons $q=p$}.    Choosing the case $n>0$ one obtains $r_e=(\sqrt{2}+1)n$ and $r_m=(\sqrt{2}-1)n$, so  $r_m<r_e$, so there are five different patterns, presented in  Fig.\ref{BH dyon q=p rh>re rh>rm}-\ref{BH dyon q=p rh<re rh<rm}.
\vspace{-.1cm}
    \vspace{-.7cm}
	\vspace{-.5cm}
    \vspace{-.5cm}
    \vspace{-.6cm}
\vspace{-.3cm}
	 {\bf Kerr-Newman-NUT}. The main new feature is that the charge densities at the horizon \eqref{kerr charge distr on horizon} can also change sign (polarization), so the HH hair can arise.
    \vspace{-.2cm}
	\begin{figure}[H]
		\centering
\includegraphics[width=0.7\textwidth]{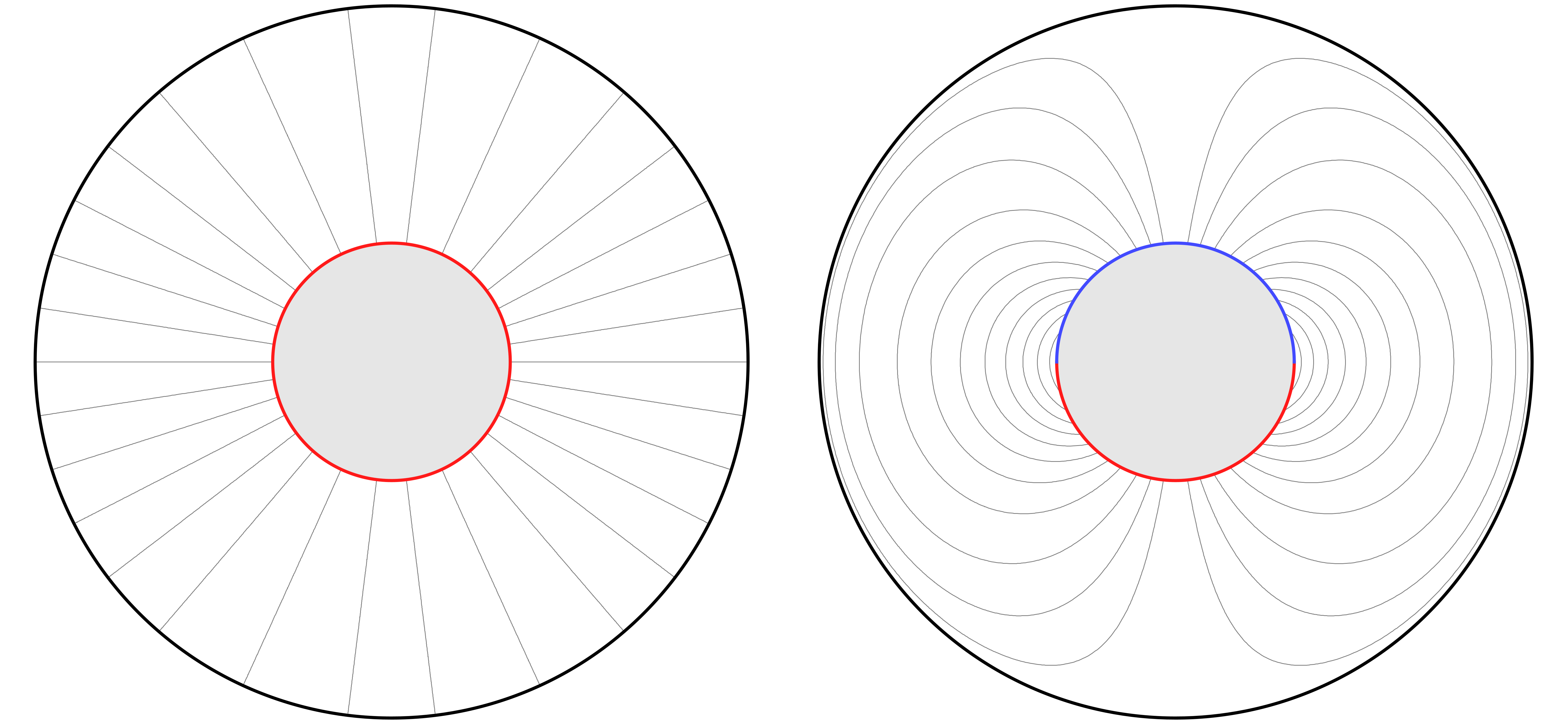 }
		\caption{Nutless electrically charged BH $m=2,\; n=0,\; q=1.32,\;p=0,\; a=-0.5$. Magnetic charge density  on the horizon, induced by rotation, has opposite signs on the northern and southern hemispheres, between which closed LFs form HH magnetic hair.}
		\label{Kerr 0}
	\end{figure}
	\begin{figure}[H]
		\centering
\includegraphics[width=0.7\textwidth]{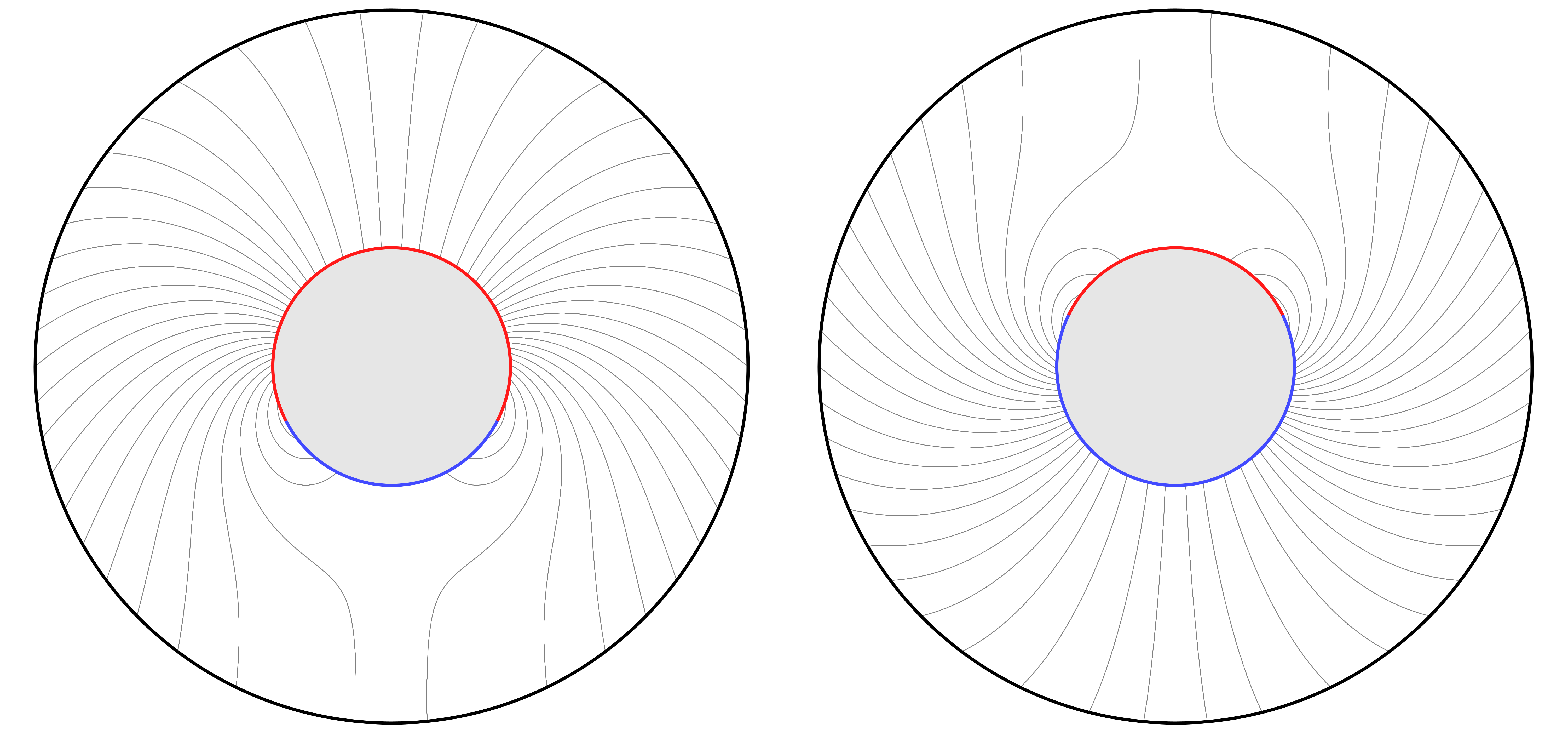}
		\caption{ NUTless $p=-q$ rotating dyon ($m=2,\; n=0,\; q=0.1,\;p=-0.1,\; a=1.99$). Both electric and magnetic horizons charge densities have positive and negative sectors, between which HH-hair arise.}
		\label{Kerr 0 1}
	\end{figure}
	\begin{figure}[H]
		\centering
		\includegraphics[width=0.7\textwidth]{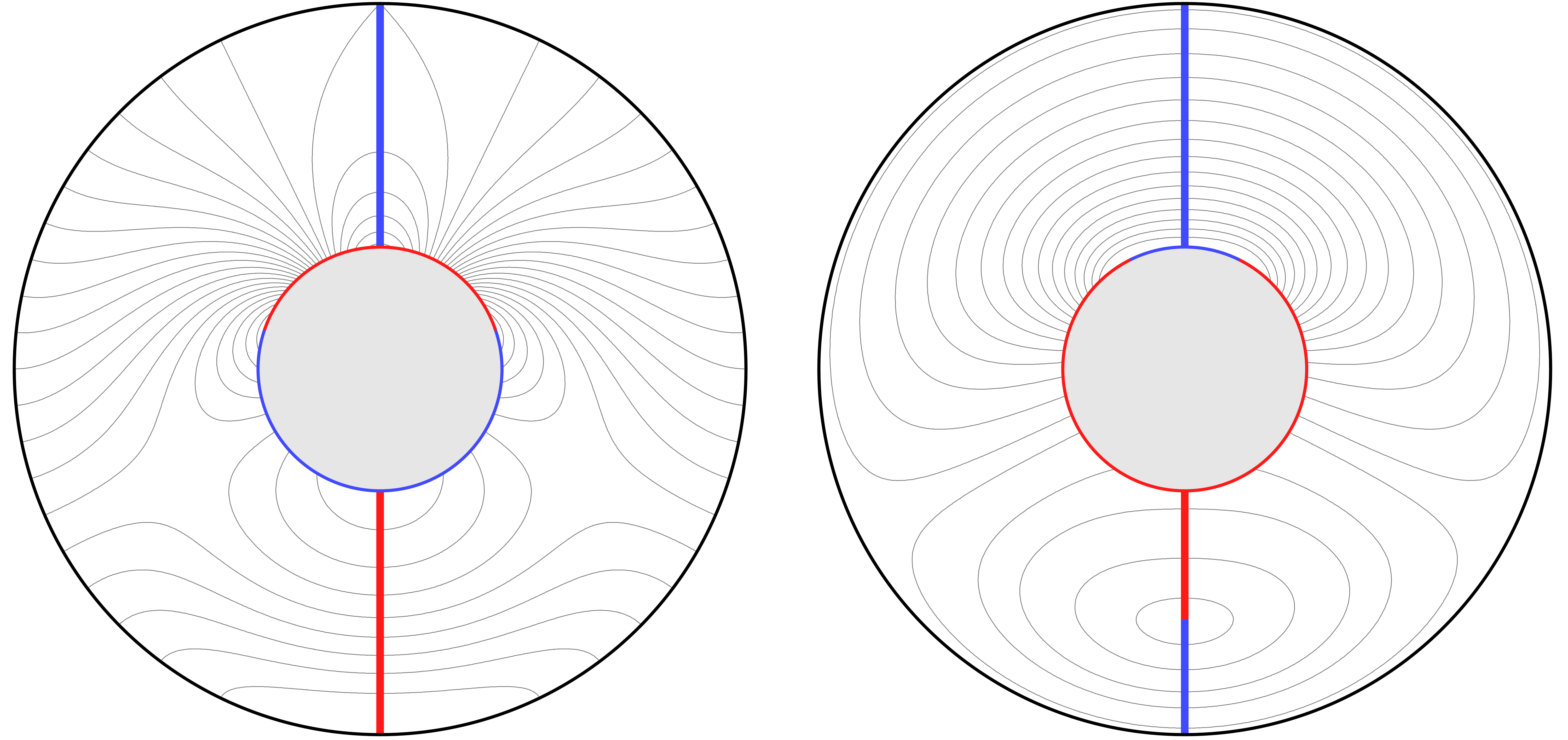 }
		\caption{Electric rotating BH with NUT ($m=0.5,\; n=1,\; q=0.01,\; p=0,\; a=-1.11$). Electric:  oppositely   charged strings and a polarized horizon, SH and HH hair. Magnetic: South string density changes sign, SS and SH hair, full confinement.}
		\label{Kerr 3}
	\end{figure}
	\begin{figure}[H]
		\centering
        \includegraphics[width=0.7\textwidth]{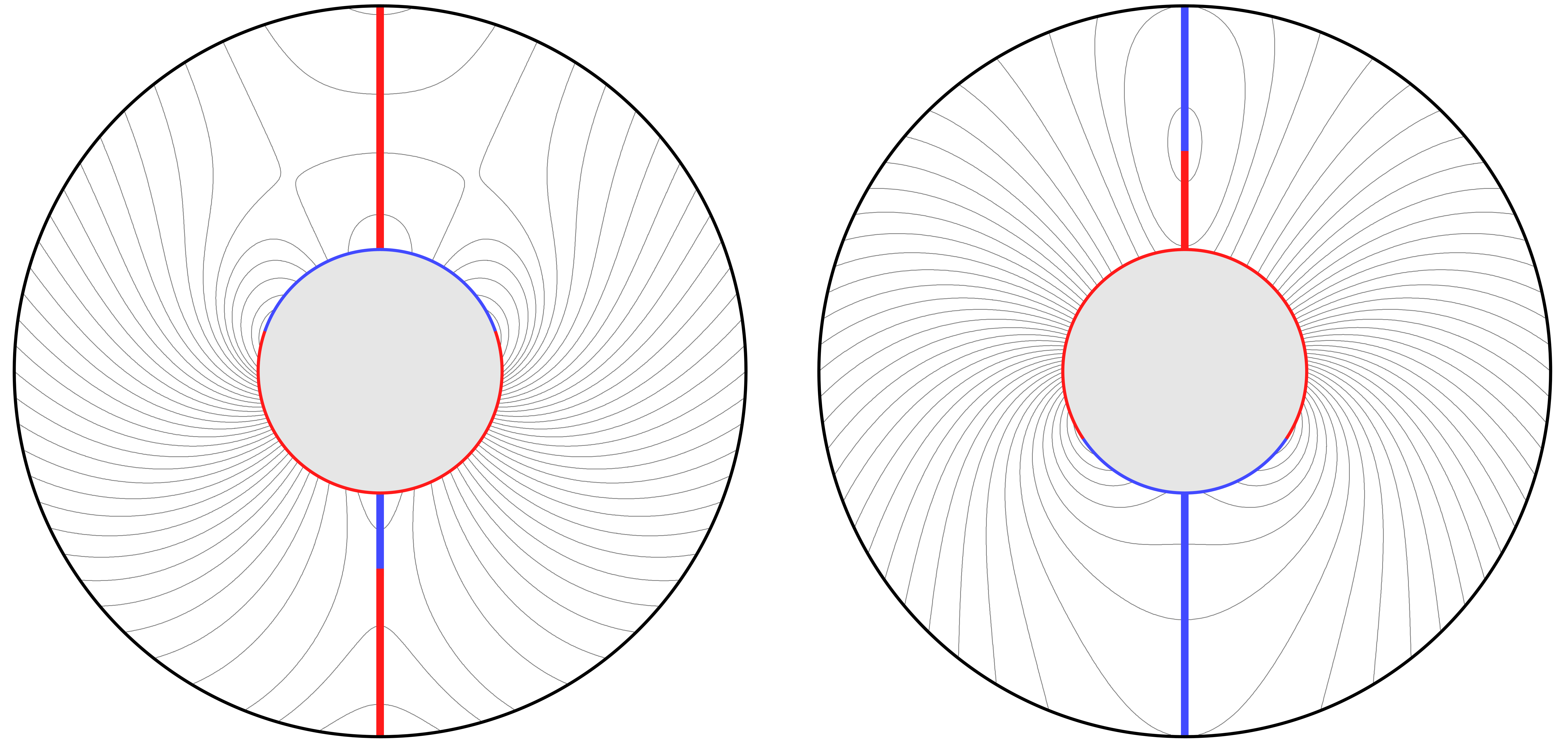 }
		\caption{Rotating symmetric dyon  ($m=0.5,\; n=0.05,\; q=p=0.07,\; a=0.492$). The most sophisticated pattern presenting all types of hair in both sectors.  }
		\label{Kerr 4}
	\end{figure}
	\section{Conclusions}
     
    Our results are based on the treatment of charged nutty solutions on the singular polar axis in terms of distributions.  Technically, this amounts to using the formula \eqref{ddphi} for the exterior derivative of the one-form $d\varphi$ that reveals the presence of the delta-like fluxes of electric and magnetic fields in the Maxwell field tensor, and similar distributional terms in the Ricci tensor\footnote{This seems to have been forgotten in two recent publications Phys. Rev. D 107, 064016, 064017 (2023), purporting to prove the fictitiousness of Dirac strings in the Einstein-Maxwell theory with both electric and magnetic  potentials. Though the mass formulas in this formulation do not include $A_\varphi$ and $B_\varphi$, these components are still present and immediately generate magnetic and electric fluxes in the Dirac strings, invalidating the claims made. More generally, such an approach is in principal incorrect since it leads to non-local theory once interactions with charged matter are included.}. Non-uniformity of the Maxwell fluxes forces the LFs escape to the bulk forming a complex electromagnetic structure which looks to be generated by effective electric and magnetic currents along the strings.

     Thus, MD strings are classically observed through a short-range electromagnetic "hair" arising from the presence of positive and negative zones of effective charge density along the strings and at the horizon. Lines of forces between them are confined and can be distinguished as SS, SH and, in the rotating case, HH hair. Our treatment is fully consistent with the discoveries of McGuire and Ruffini \cite{Mcguire:1975wq} providing further clarifications of their results. It is also consistent with previous work in this direction \cite{Clement:2017otx,Clement:2019ghi,Bordo:2019slw, BallonBordo:2020mcs,Abbasvandi:2021nyv,Awad:2022jgn} and with some results based on the membrane paradigm in black hole physics \cite{Thorne:1986iy}. We presented a simple example of similar analysis in supergravity
   and hope that our approach may  be useful in other applications.
  
{\bf Acknowledgments.} The authors thank G\'erard Cl\'ement for valuable suggestions
and discussions.  The work  was supported by the Foundation for the Advancement of Theoretical
Physics and Mathematics 'BASIS'.\newpage
	\appendix
    \renewcommand{\theequation}{A.\arabic{equation}}
    \setcounter{equation}{0}
	\section{Supergravity: Gal'tsov-Kechkin solution.}
	\label{EMDA}
Here we extend our results to supergravity, exploring the seven-parametric solution of ${\cal N}=4$ theory found by solution generating technique in \cite{Galtsov:1994pd} and recently derived by direct integration of the field equations in \cite{Galtsov:2025nia}, whose notation we are using. The calculation of the electric charge densities on the MD strings and the horizon gives
\begin{align*}
&\rho^\pm_e=n\frac{p(r^2-n_\pm^2+dd^*)+2r(q n_\pm+u)}{(r^2+n_\pm^2-dd^*)^2},\\
&\rho^H_e=\frac{q(r_+^2-\nu^2-dd^*)-2\nu(pr_++u)}{4\pi (r_+^2+\nu^2-dd^*)^2},
\end{align*}
where $n_\pm=n\pm a,\;\nu=n+a\cos\theta,\; u={\rm Im}[d(q+ip)]$ and $d=(q-ip)^2/[2(m+in)]$ is the axidilaton charge.  Magnetic charge densities  $\rho^\pm_m$ and $\rho^H_m$ are obtained by the same duality substitution as before : $q\rightarrow p,\;p\rightarrow-q$, also leading to   interchange ${\rm Im}[d(q+ip)]\leftrightarrow {\rm Re}[d(q+ip)]$.

Note that one can recover the charge densities for the Kerr-Newman-NUT solution \eqref{Kerr strings}, \eqref{kerr charge distr on horizon} by formally setting $d=0$, which is, of course, illegal, since axidilaton charge depends on other parameters.  Since $d$ depends on the mass $m$, the charge densities become dependent on it as well, in contrast to the Einstein-Maxwell case. Another new feature is that both roots of the equations $\rho^\pm_e(r)=0$ and $\rho^\pm_m(r)=0$ can now be  positive at some values of the parameters. 

\begin{figure}[H]
	\centering
\includegraphics[width=0.7\textwidth]{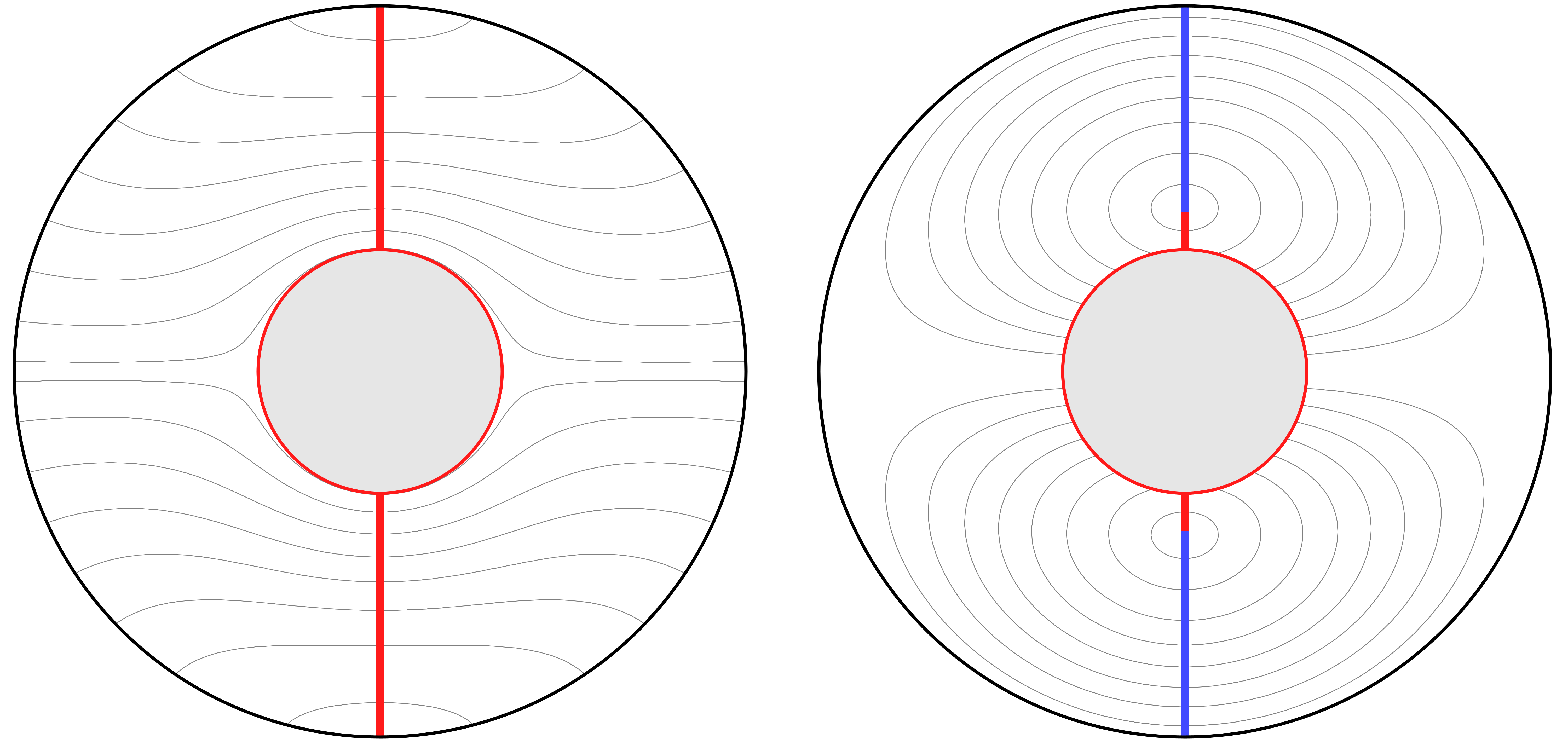}
	\caption{ $m=0.15,\; n=2,\; q=1.81,\; p=0.1,\; a=0$. No electric separatrix and all the electric LFs propagate to infinity, though the MD strings  magnetic charge densities change sign, so all magnetic LFs are confined (SS and SH hair).}
	\label{EMDA 1}
\end{figure}

\begin{figure}[H]
	\centering
	\includegraphics[width=0.7\textwidth]{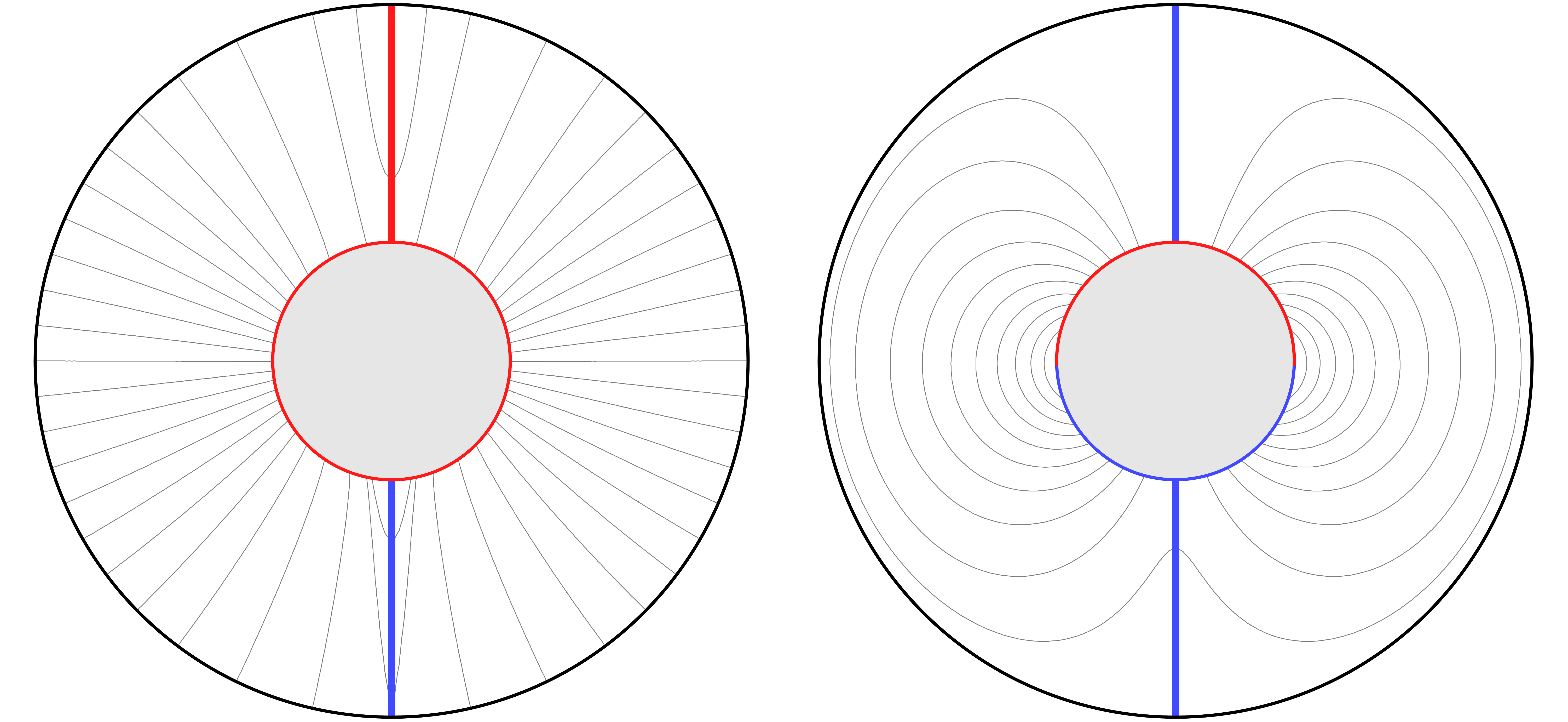}
	\caption{ $m=0.44,\; n=0.01,\; q=0.2,\; p=0,\; a=0.343$.  Electric panel: oppositely  charged strings and positive horizon. Magnetic charge density of the horizon changes sign, so both SH and HH hair are present.}
	\label{EMDA 2}
\end{figure}

\begin{figure}[H]
		\centering
\includegraphics[width=0.7\textwidth]{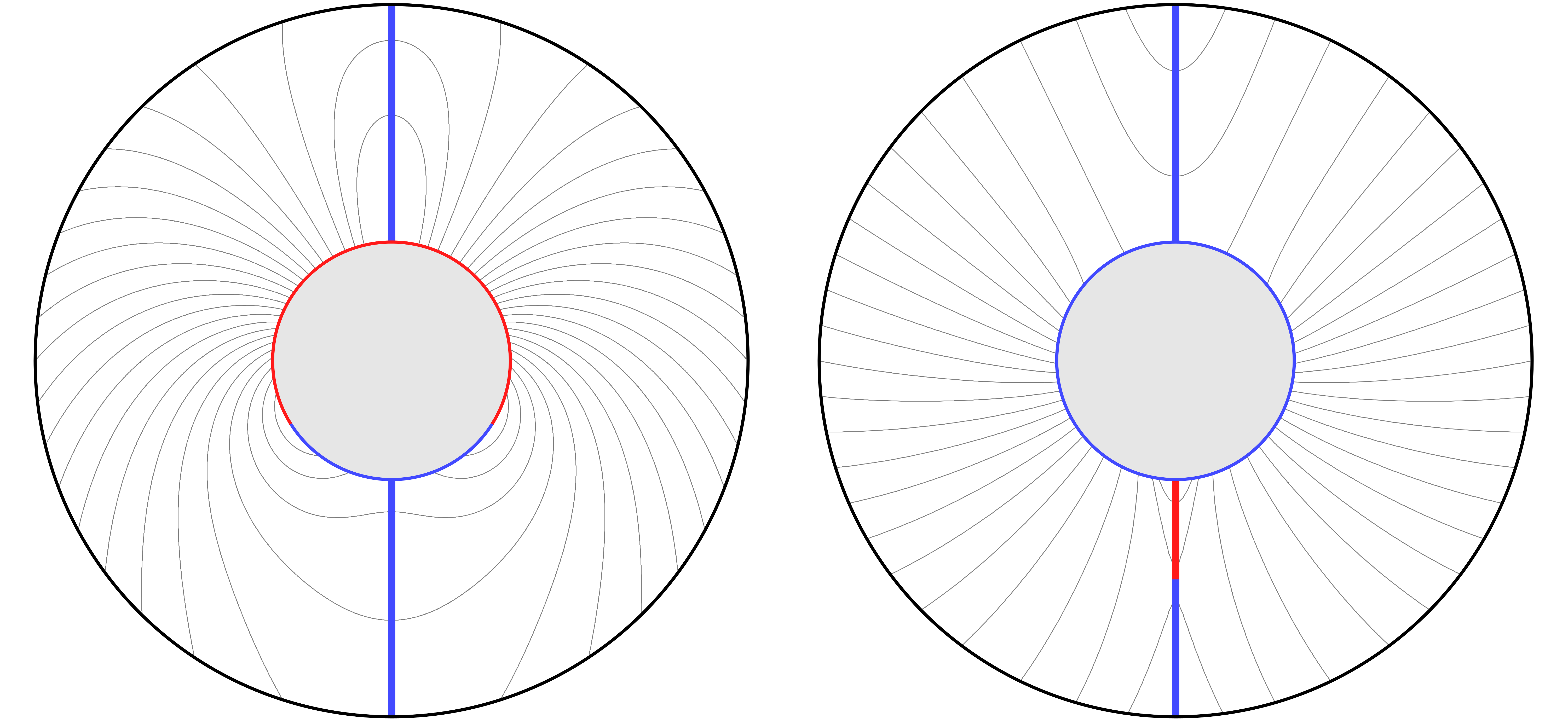 }
		\caption{ $m=2,\; n=0.16,\; q=0.42,\; p=-1,\; a=1.5$.  SH and HH electric hair,   SH magnetic hair.}
		\label{EMDA 6}
	\end{figure}

\begin{figure}[H] 
		\centering
\includegraphics[width=0.7\textwidth]{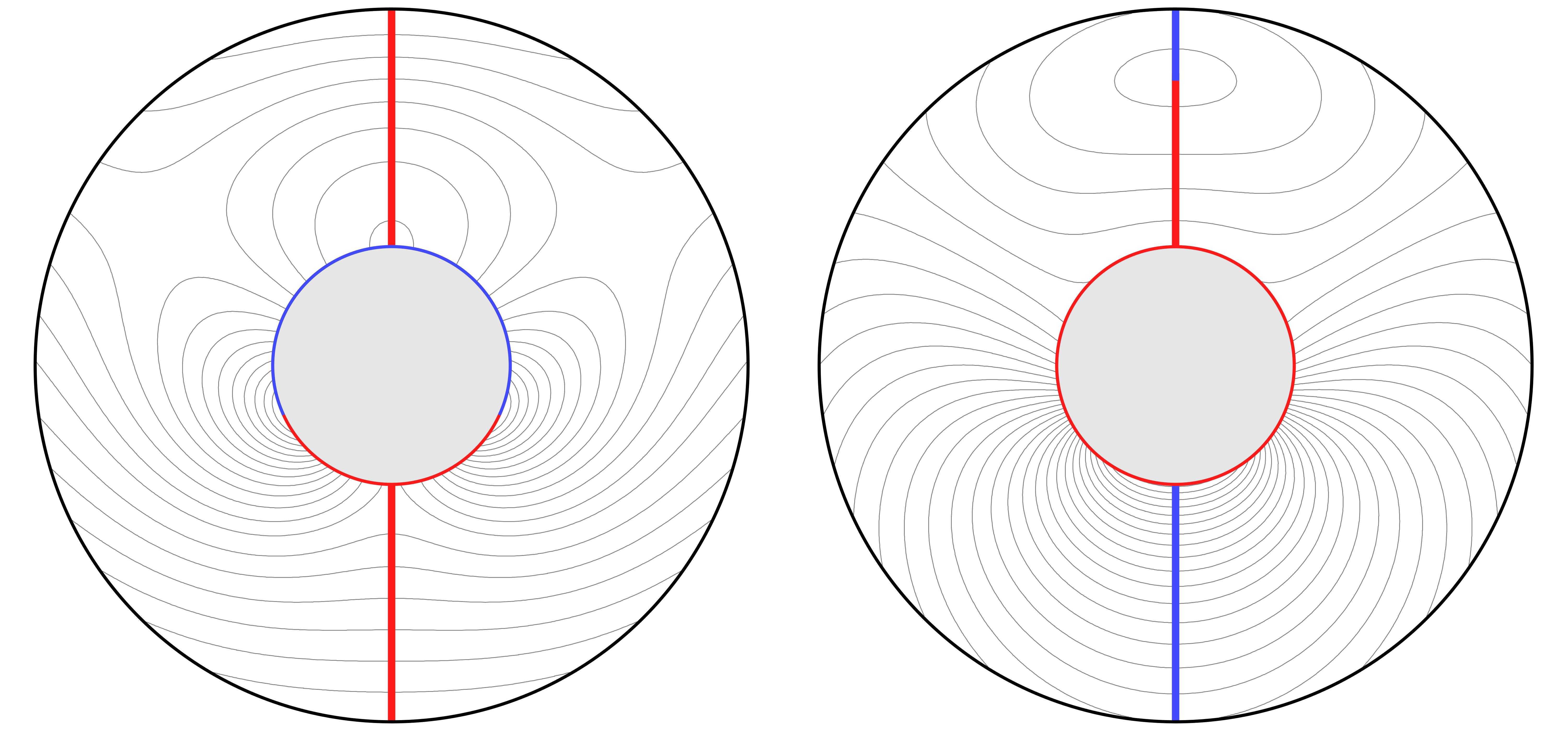  }
		\caption{ $m=0.5,\; n=0.5,\; q=0.07,\; p=0.07,\; a=0.7$. Electric HH hair and  both SS and SH hair.}
		\label{EMDA 7}
	\end{figure}

{\bf Static solution.} One can see, that in the case of vanishing rotation the transition points $r_e$ and $r_m$ no longer coincide with the points where the MD strings  charge densities change signs Fig.\ref{EMDA 1}.
\vspace{0.3 cm} 
\newpage
{\bf Rotating dyon.} Rotating  electric solution without NUT produces exactly the same LF pattern as the Kerr-Newman case shown in Fig. \ref{Kerr 0}. Its deformation with a small NUT parameter is shown in Fig. \ref{EMDA 2}. Varying other parameters produces different hair patterns:  Fig.\ref{EMDA 6}-Fig.\ref{EMDA 7}.

\end{document}